# Agreement Between Large Language Models, Human Reviewers, and Authors in Evaluating STROBE Checklists for Observational Studies in Rheumatology

*This manuscript is a preprint and has not undergone peer review.*

Emre Bilgin[1], Ebru Öztürk[2], Meera Shah[3], Lisa Traboco[4,5], Rebecca Everitt[6], Ai Lyn Tan[7], Marwan Bukhari[8,9], Vincenzo Venerito*[10], Latika Gupta*[11,12,13]

[1] Sakarya University, Faculty of Medicine, Department of Internal Medicine, Division of Rheumatology, Sakarya, Turkiye
[2] Hacettepe University, Faculty of Medicine, Department of Biostatistics, Ankara, Turkiye
[3] Topiwala National Medical College and B.Y.L. Nair Hospital, Department of Rheumatology, , Mumbai, India
[4] St Luke's Medical Center (Global City), Philippines
[5] University of the Philippines (Medical Informatics Unit), Manila, Philippines
[6] New Cross Hospital, The Royal Wolverhampton Trust, Wolverhampton, United Kingdom
[7] NIHR Leeds Biomedical Research Centre, Leeds Teaching Hospitals NHS Trust, Leeds, UK; Leeds Institute of Rheumatic and Musculoskeletal Medicine, University of Leeds, Leeds, United Kingdom
[8] Lancaster University Medical School, Lancaster University, Lancaster, United Kingdom
[9] Department of Rheumatology, Royal Lancaster Infirmary, Lancaster, United Kingdom
[10] Rheumatology Unit, Department of Precision and Regenerative Medicine and Ionian Area (DiMePReJ), University of Bari "Aldo Moro", Piazza Umberto I, 70121, Bari, Italy
[11] Department of Rheumatology, Royal Wolverhampton Hospitals NHS Trust, Wolverhampton, UK.
[12] School of Infection, Inflammation and Immunology, College of Medicine and Health, University of Birmingham, Birmingham, UK.
[13] Francis Crick Institute, London, UK.
*Co-senior authors

ORCids:

Emre Bilgin: 0000-0002-2260-4660
Ebru Öztürk: 0000-0001-9206-6856
Meera Shah: 0009-0005-3308-7231
Lisa Traboco: 0000-0002-1952-7879
Rebecca Everitt: 0000-0003-1919-5293
Ai Lyn Tan: 0000-0002-9158-7243
Marwan Bukhari: 0000-0003-4311-5222
Vincenzo Venerito: 0000-0002-2573-5930
Latika Gupta: 0000-0003-2753-2990

**Correspondence:**
Vincenzo Venerito
Rheumatology Unit, Department of Precision and Regenerative Medicine and Ionian Area (DiMePReJ),
University of Bari "Aldo Moro",
Piazza Umberto I, 70121, Bari, Italy
vincenzo.venerito@gmail.com

**Key messages:**

- In our sample, LLMs showed high agreement with human reviewers on basic STROBE formatting, indicating a potential role in initial screening.
- However, LLM evaluations of complex methodological items may differ from those of senior human experts.
- Although models may facilitate structural checks, expert human judgment remains crucial for evaluating methodology.

**ABSTRACT**

**Introduction:** Evaluating compliance with the Strengthening the Reporting of Observational studies in Epidemiology (STROBE) statement can be time-consuming and subjective. This study compares STROBE assessments from large language models (LLMs), a human reviewer panel, and the original manuscript authors in observational rheumatology research.

**Methods:** Guided by the Guidelines for Reporting Reliability and Agreement Studies (GRRAS) and DEAL Pathway B frameworks, 17 rheumatology articles were independently assessed. Evaluations used the 22-item STROBE checklist, completed by the authors, a five-person human panel (ranging from junior to senior professionals), and two LLMs (ChatGPT-5.2, Gemini-3Pro). Items were grouped into "Methodological Rigor" and "Presentation&Context" domains. Inter-rater reliability was calculated using Gwet's Agreement Coefficient (AC1).

**Results:** Overall agreement across all reviewers was 85.0% (AC1=0.826). Domain stratification showed almost perfect agreement for "Presentation&Context" (AC1=0.841) and substantial agreement for "Methodological Rigor" (AC1=0.803). Although LLMs achieved complete agreement (AC1=1.000) with all human reviewers on standard formatting elements, their agreement with human reviewers and authors declined on complex items. For example, regarding the item on loss to follow-up, the agreement between Gemini 3 Pro and the senior reviewer was AC1=-0.252, while the agreement with the authors was only fair. Additionally, ChatGPT-5.2 generally demonstrated higher agreement with human reviewers than Gemini-3Pro on specific methodological items.

**Conclusion:** While LLMs show potential for basic STROBE screening, their lower agreement with human experts on complex methodological items likely reflects a reliance on surface-level information. Currently, these models appear more reliable for standardizing straightforward checks than for replacing expert human judgment in evaluating observational research.

**INTRODUCTION**

Since the introduction of earlier models such as Generative Pre-trained Transformer 3 (GPT-3), large language models (LLMs) have quickly become a part of academic research [1]. They are increasingly used for tasks such as literature screening, data extraction, and code generation, helping to reduce manual workload and human error [2]. However, their use in critical scientific processes is still debated, mainly because of unresolved issues related to reliability, bias, and transparency. [3].

Transparent reporting is crucial for assessing observational research. The Strengthening the Reporting of Observational Studies in Epidemiology (STROBE) Statement provides a 22-item checklist to standardize this process and improve interpretability [4]. In practice, however, evaluating STROBE adherence is time-consuming and subject to inter-rater variability because it lacks strictly objective criteria [5]. Furthermore, as Sharp et al. noted, many authors view the checklist as a time-consuming administrative requirement rather than a practical tool, which can compromise the accuracy of self-reported adherence [5].

In this study, we aimed to systematically evaluate how well current LLMs can assess adherence to the STROBE checklist in observational rheumatology studies. By directly comparing their performance with a panel of human reviewers who have different levels of clinical expertise, as well as with the original manuscript authors, we sought to explore the potential of artificial intelligence (AI) to serve as a supplementary tool for standardizing reporting assessments and helping readers and researchers in the critical appraisal of observational literature.

**METHODS**

The study design, inter-rater agreement analyses, and reporting framework were guided by the principles of the Guidelines for Reporting Reliability and Agreement Studies (GRRAS) and aligned with the Development, Evaluation, and Assessment of Large Language Models (DEAL) Pathway B checklist for the applied evaluation of pre-trained models [6, 7].

*Search Strategy and Selection Criteria*:

A list of top rheumatology journals, ranked by Impact Factor, was compiled using the Observatory of International Research (OOIR) [8]. Journal-specific STROBE requirements were verified via prior bibliometric analyses and the respective "Guide for Authors" guidelines [9, 10]. Within the shortlisted journals, a structured search identified all articles that mentioned "STROBE" published between February 2020 and February 2025. 17 articles containing complete, accessible STROBE checklists were included for evaluation (for full references, **Supplementary Text-1**). The selection process is detailed in the study selection flowchart (**Supplementary Figure-1**).

***Evaluation Protocol*:**

A panel of human reviewers—comprising three junior, one mid-level, and one senior professional—assessed the 17 articles independently using the STROBE checklist. Additionally, two LLMs, ChatGPT-5.2 and Gemini 3 Pro, conducted assessments. For comparison, the authors' original, self-reported STROBE checklists were also collected.

To ensure reproducibility and adhere to the DEAL-B framework, evaluations used the premium subscription levels of two widely used, proprietary, closed-source models via their official web interfaces: ChatGPT-5.2 (Auto mode; OpenAI) and Gemini 3 Pro (Google LLC). All tests were performed on January 10, 2026. Web interfaces were chosen over APIs to simulate real-world clinical settings, with generation parameters like temperature and top-p maintained at vendor default settings. Each manuscript was assessed in a new chat session to ensure independent evaluations. We employed a three-step prompting method (full prompt texts available in **Supplementary Text-2**). This included: (1) instructing the models to evaluate only reporting adherence rather than methodological quality, using strict Y/N/NA categories, and advising them not to infer missing details; (2) a few-shot calibration step with two example articles with human reference scores to set the threshold for adequate reporting; and (3) the final evaluation, which involved a simple two-column table without explanations or reasoning to prevent the models from generating fabricated explanations. All model outputs were recorded exactly as they were generated, without human modification.

***Scoring and Statistical Analysis:***

All reviewers (human and machine) categorized each STROBE item as "Yes" (adequately reported), "No" (inadequately or not reported), or "Not applicable." Before beginning data collection, our research team grouped STROBE items into two categories: 'Methodological Rigor'

(Items 4–12) and 'Presentation & Context' (Items 1–3, 13–22). This matches the original STROBE checklist, in which items 4–12 constitute the 'Methods' section. We planned this grouping to see if the models perform differently when evaluating core methodological steps and general reporting practices. We assessed inter-rater reliability rather than diagnostic accuracy (e.g., sensitivity), because evaluating reporting guidelines inherently relies on subjective interpretation and lacks an absolute 'ground truth'. Inter-rater reliability across human reviewers, LLMs, and original manuscript authors was quantified using overall agreement proportions and Gwet's First-Order Agreement Coefficient (AC1) with 95% confidence intervals (CIs) by using ''irrCAC'' package [11]. Gwet's AC1 was selected to estimate reliability, as it corrects for chance agreement while avoiding the prevalence paradox frequently encountered with Cohen's kappa in highly skewed checklist datasets. Reliability estimates were interpreted using standard thresholds: <0.00 (poor), 0.00–0.20 (slight), 0.21–0.40 (fair), 0.41–0.60 (moderate), 0.61–0.80 (substantial), and 0.81–1.00 (almost perfect) [12]. All statistical analyses to evaluate overall, domain-specific, and item-specific agreement across all evaluator configurations were conducted using R software version 4.4.3 (R Foundation for Statistical Computing, Vienna, Austria).

**RESULTS**

**Overall and Domain-Specific Agreement**

Of the 189 screened manuscripts, only 17 articles provided a complete and accessible STROBE checklist suitable for analysis (**Supplementary Figure 1**). Overall agreement across all items and reviewers (human, LLMs, authors) was 85.0% (Gwet's AC1 = 0.826 [95% CI 0.801–0.851]), indicating almost perfect agreement. Stratification by domain revealed almost perfect agreement for "Presentation & Context" (AC1 = 0.841 [95% CI 0.810–0.872]) and substantial agreement for "Methodological Rigor" (AC1 = 0.803 [95% CI 0.761–0.845]), with all p-values < 0.001. Pairwise analysis demonstrated highly consistent inter-rater reliability among all evaluation groups, with the vast majority of combinations exhibiting almost perfect agreement (AC1 > 0.80; Figures 1 and 2). At the article level, overall AC1 ranged from 0.624 to 0.931, with greater variance observed in the "Methodological Rigor" domain (0.532–0.921) compared to "Presentation & Context" (0.685–0.946).

**Item-Level Analysis**

An examination of the agreement scores for individual STROBE items and sub-items revealed that while reviewers reached near-perfect agreement on many components, concordance naturally varied across the checklist. Noticeable differences—and in some cases, considerably lower scores—emerged on specific questions. Detailed pairwise comparisons for each specific item across all evaluator groups are provided in Supplementary Figure-2.

- **High Agreement on Standard Reporting Items**

Within the "Presentation & Context" domain, both language models consistently identified standard manuscript sections. For basic reporting elements—such as the informative abstract (Item 1b), background and rationale (Item 2), specific objectives (Item 3), summary of key results (Item 18), study limitations (Item 19), and overall interpretation (Item 20)—ChatGPT-5.2 and Gemini 3 Pro showed perfect agreement (AC1 = 1.000) with all human reviewers and the original authors. Agreement was similarly high for identifying funding sources (Item 22) and participant numbers (Item 14a).

- **Discrepancies in Complex Methodological Items**

Item-level analysis showed greater variation, especially in methodological reporting, specifically in Item 12. Agreement between the LLMs and the senior reviewer declined on more complex items. For example, regarding loss to follow-up (Item 12d), LLM-to-LLM agreement was moderate (AC1 = 0.447), and both models showed fair agreement with the authors and junior reviewers. However, their agreement with the senior reviewer dropped significantly; the agreement between Gemini 3 Pro and the senior reviewer fell within the negative range (AC1 = -0.252). For sensitivity analyses (Item 12e), LLMs agreed substantially with each other (AC1 = 0.652) and moderately with the senior reviewer but showed slight to fair agreement with the authors (AC1: 0.071–0.272). Finally, regarding reporting continuous-variable categorizations (Item 16b), LLMs agreed moderately with one another (AC1 = 0.564), and human groups largely agreed among themselves; however, agreement between LLMs and human reviewers fell into the fair and slight categories.

- **Performance Variations Between Language Models**

Agreement levels also differed between the two language models themselves. When assessing how missing data was handled (Item 12c), despite substantial LLM-to-LLM agreement (AC1 = 0.668), junior reviewers showed low concordance with both models, and Gemini 3 Pro generally exhibited lower agreement with human reviewers than ChatGPT-5.2. The authors maintained substantial agreement with junior reviewers on this item. Model performance varied regarding reporting the number of participants with missing data (Item 14b). ChatGPT-5.2 achieved substantial agreement with both authors (AC1 = 0.730) and the mid-level reviewer (AC1 = 0.722), whereas Gemini 3 Pro demonstrated fair (AC1 = 0.301) and moderate (AC1 = 0.450) agreement.

**DISCUSSION**

In this study, we examined the level of agreement between large language models, a panel of human reviewers, and manuscript authors in assessing adherence to the STROBE checklist in observational rheumatology research. Although the overall agreement was 85.0%, this was primarily driven by the "Presentation & Context" domain. Items in this section generally require identifying standard structural components, which aligns well with the basic text-matching nature of current language models [13]. In contrast, agreement was lower and more variable in the "Methodological Rigor" domain, suggesting that evaluating complex methodological reporting remains challenging for all types of reviewers.

The item-level results suggest that a reviewer's background likely influences their scoring patterns. Regarding the reporting of sensitivity analyses (Item 12e), authors generally agreed with junior reviewers but showed only slight to fair agreement with the LLMs. The models' scores were more aligned with those of the senior reviewer. This finding is consistent with existing research indicating that authors tend to view their own methodological reports more positively than independent reviewers do [14]. In our sample, junior reviewers appeared more inclined to accept authors' statements, whereas senior experts and language models tended to apply stricter standards. Consequently, these findings suggest that incorporating LLMs might offer a useful supplementary check for verifying author-reported checklists.

We also observed instances in which model assessments diverged from human-reviewer judgments. For tasks such as handling loss to follow-up (Item 12d) and categorizing continuous variables (Item 16b), the agreement between the models and the senior reviewer fell to slight or poor levels. This likely happens because the models rely heavily on literal wording. If a paper does not clearly specify how loss to follow-up was managed, the model tends to consider it as missing data. In contrast, an experienced human reviewer can often infer this information from data tables or flowcharts. This is a recognized limitation: models generally perform well with surface-level text but struggle to recognize implicit methodological details [15]. Future studies should include qualitative analyses of these discordant cases to better understand the specific reasons behind these disagreements.

Apart from differences between human and model outputs, performance gaps were observed between the two language models. For example, in assessing missing data counts (Item 14b), ChatGPT-5.2 closely aligned with human reviewers, while Gemini 3 Pro had lower agreement scores. Prior research on medical benchmarks similarly indicates that outputs can vary with models' training and update cycles [16]. This underscores the importance of researchers carefully choosing their models, as the selected tool can significantly impact evaluation results.

Our study has several key strengths. First, we assembled a diverse panel of human reviewers, including both junior researchers and senior experts, alongside the original manuscript authors. This setup enabled us to directly observe how clinical experience influences checklist scoring. Second, we tested two language models using a standardized, isolated-prompting protocol without human intervention, ensuring that our machine evaluations are reproducible. Third, we used Gwet's AC1 statistic to measure agreement, which avoids the prevalence paradox often seen with Cohen's kappa in checklist data, yielding a more accurate estimate of inter-rater reliability. However, our study also has limitations. The primary limitation is the small sample size of 17 articles, which naturally reduces statistical power and widens confidence intervals, particularly for item-level analyses. However, this number was not a methodological choice but a direct reflection of the current literature. Despite screening a large number of observational studies in rheumatology, we found only 17 that explicitly reported using the STROBE checklist. While this small sample makes our estimates less stable, it also points to a broader, critical issue:

widespread lack of reporting of this guideline in our field. Also, since our sample consists of articles from high-impact journals that explicitly mandate STROBE checklists, the reporting quality we evaluated is likely higher than average. Across a broader set of publications, the agreement between LLMs and human reviewers might differ. Analytically, grouping the STROBE checklist into two domains helped us separate methods from formatting. We recognize this split is imperfect, though; for example, Item 16b falls within the presentation domain, but its low agreement scores indicate it behaves more like a complex methodological task. We also evaluated only two commercial LLMs, excluding open-source models such as Llama and Mistral. We made this choice to reflect the daily reality of researchers, who generally rely on standard, ready-to-use web interfaces rather than on local installations that require specific hardware. Future studies should explore how open-source or specialized medical models perform in these tasks. We also chose to evaluate each article only once per model, meaning we did not assess intra-rater consistency. Language models can generate different responses to the same prompt, and multiple runs can show how much their answers fluctuate. However, we intentionally used a single-pass approach to mirror the real-world workflow of human reviewers, who typically read and score a manuscript just once. Allowing multiple runs to select the most accurate outputs would have also introduced selection bias. For future applications, integrating Retrieval-Augmented Generation (RAG) systems offers a practical solution to address this natural variability [17]. By anchoring the model to specifically retrieved text segments, RAG architectures can make automated checklist evaluations significantly more stable and reliable [17]. There is also an inherent risk of data leakage. Because the evaluated articles are already public, they are likely part of the models' training data. The models might have relied on prior exposure rather than extracting information solely from the texts we provided, which could skew the agreement rates upwards. Even with strict prompting to use only the provided text, it is impossible to separate an LLM from its pre-training memory. Beyond this pre-training memory, the models themselves are constantly evolving; because we tested ChatGPT-5.2 and Gemini 3 Pro at a specific point in time, future updates may naturally yield different results.

In conclusion, language models show potential as a practical tool for assessing general reporting standards, making them useful for basic STROBE screening. However, their agreement with senior human experts declines on complex methodological items, likely reflecting a reliance on surface-level information rather than a deeper methodological understanding. While these models can help standardize straightforward checklist tasks, expert human judgment should remain essential for accurately evaluating study methods in observational research.

*Supplementary material is provided at the end of this manuscript.*

**Ethics Approval**

Ethical approval was not required for this study, as it exclusively involved the methodological analysis of previously published, publicly available observational studies and did not involve human or animal subjects.

**Data Availability Statement**

The datasets generated and analyzed during the current study, including the de-identified scoring sheets from all human raters and the complete verbatim transcripts of the language model outputs, are available from the corresponding author upon reasonable request.

**Funding**

The authors received no specific grant from any funding agency in the public, commercial, or not-for-profit sectors for the research, authorship, and/or publication of this article.

**Artificial Intelligence (AI) Disclosure**

During the preparation of this manuscript, the authors used Gemini 3 Pro (Google) solely for language editing, grammatical refinement, and readability improvement. After using this tool, all authors thoroughly reviewed, edited, and approved the final manuscript. The authors take full and final responsibility for the integrity, accuracy, and original content of this publication.

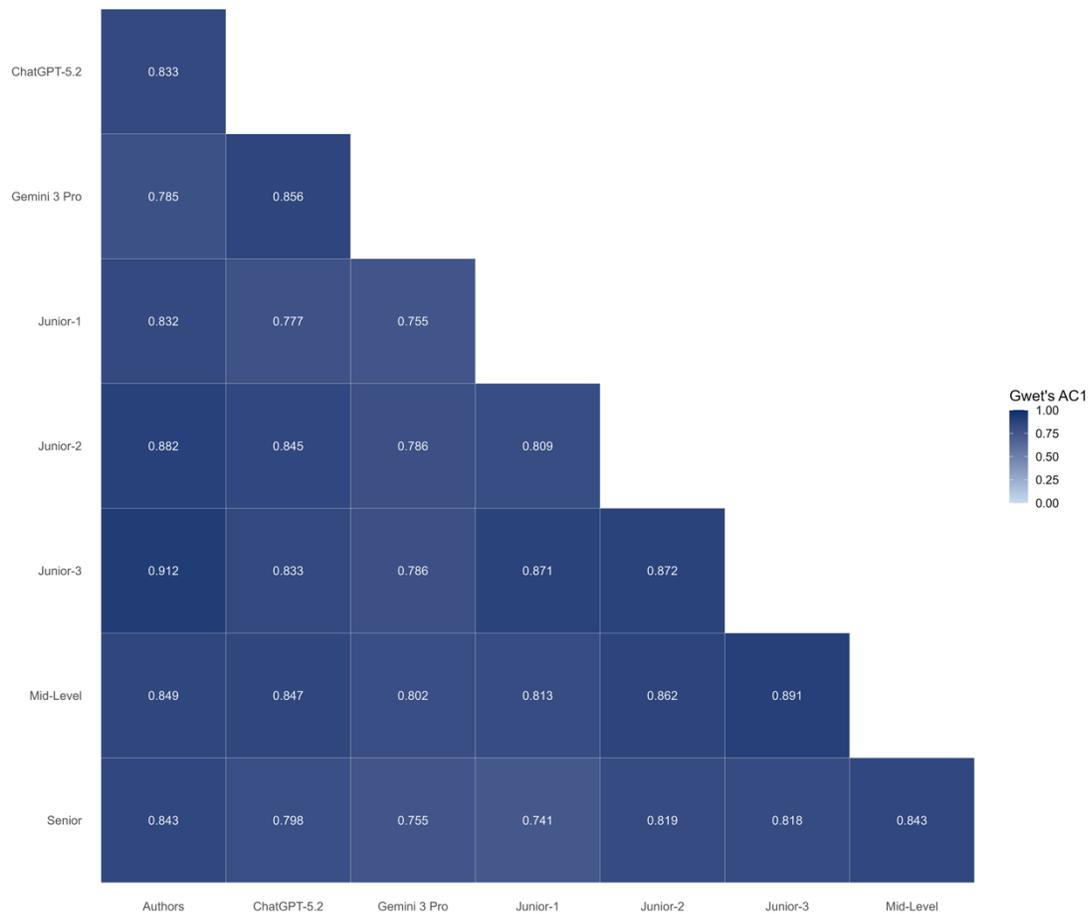

Figure 1. **Overall Pairwise Agreement Among Reviewers.** Heatmap displaying Gwet's First-Order Agreement Coefficient (AC1) estimates for overall pairwise comparisons across all STROBE checklist items. Reviewers include the original manuscript authors, two large language models (ChatGPT-5.2 and Gemini 3 Pro), and a panel of independent human reviewers (three juniors, one mid-level, and one senior).

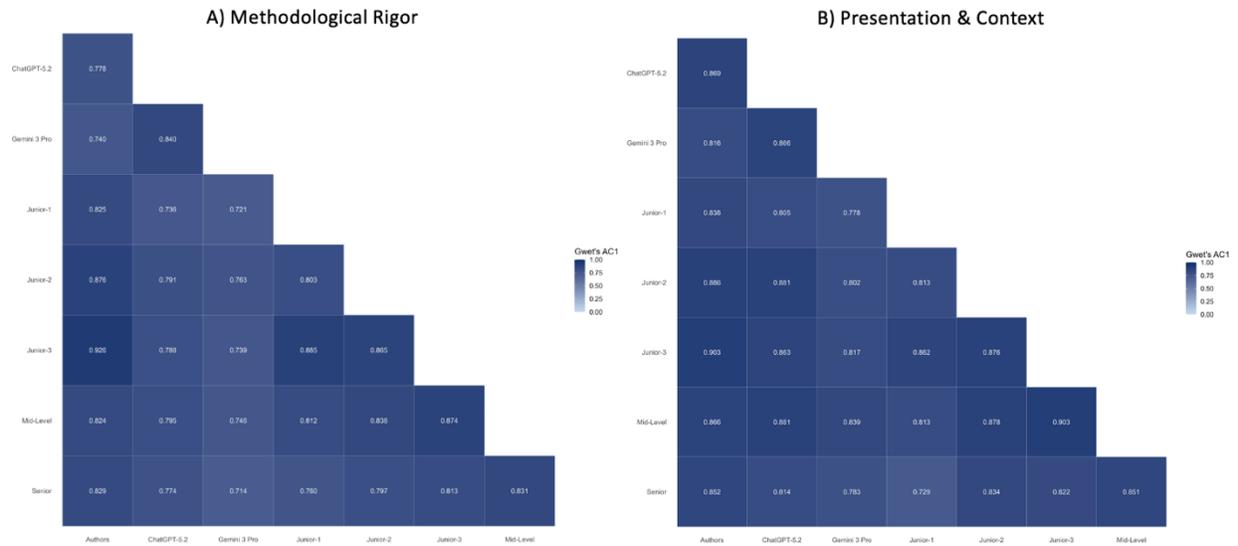

**Figure 2. Pairwise Agreement Stratified by STROBE Domains.** Heatmaps displaying Gwet's First-Order Agreement Coefficient (AC1) estimates for pairwise comparisons across two distinct evaluation domains: (A) Methodological Rigor (Items 4–12) and (B) Presentation & Context (Items 1–3, 13–22). Reviewers include the original manuscript authors, two large language models (ChatGPT-5.2 and Gemini 3 Pro), and a panel of independent human reviewers (three juniors, one mid-level, and one senior).

REFERENCES:

1. Venerito, V. and L. Gupta, *Large language models: rheumatologists' newest colleagues?* Nat Rev Rheumatol, 2024. **20**(2): p. 75-76.
2. Venerito, V., et al., *AI am a rheumatologist: a practical primer to large language models for rheumatologists.* Rheumatology (Oxford), 2023. **62**(10): p. 3256-3260.
3. Venerito, V., et al., *Ethical challenges and regulatory pathways for artificial intelligence in rheumatology.* Rheumatol Adv Pract, 2025. **9**(2): p. rkaf035.
4. von Elm, E., et al., *The Strengthening the Reporting of Observational Studies in Epidemiology (STROBE) statement: guidelines for reporting observational studies.* J Clin Epidemiol, 2008. **61**(4): p. 344-9.
5. Sharp, M.K., K. Glonti, and D. Hren, *Online survey about the STROBE statement highlighted diverging views about its content, purpose, and value.* J Clin Epidemiol, 2020. **123**: p. 100-106.
6. Kottner, J., et al., *Guidelines for Reporting Reliability and Agreement Studies (GRRAS) were proposed.* J Clin Epidemiol, 2011. **64**(1): p. 96-106.
7. Tripathi, S., et al., *Development, Evaluation, and Assessment of Large Language Models (DEAL) Checklist: A Technical Report.* NEJM AI, 2025. **2**(6): p. AIp2401106.
8. Observatory, et al.; Available from: https://ooir.org/.
9. Sharp, M.K., et al., *A cross-sectional bibliometric study showed suboptimal journal endorsement rates of STROBE and its extensions.* J Clin Epidemiol, 2019. **107**: p. 42-50.
10. Barajas-Ochoa, A., et al., *The Use of Reporting Guidelines in Rheumatology: A Cross-Sectional Study of Over 850 Manuscripts Published in 5 Major Rheumatology Journals.* J Rheumatol, 2023. **50**(7): p. 939-943.
11. Gwet, K.L. *irrCAC: Computing Chance-Corrected Agreement Coefficients (CAC).* 2019; Available from: https://CRAN.R-project.org/package=irrCAC.
12. Gwet, K.L., *Handbook of inter-rater reliability : the definitive guide to measuring the extent of agreement among raters*. 2014, Maryland, United States: Gaithersburg, MD : Advanced Analytics, LLC.
13. Garcia-Carmona, A.M., et al., *Leveraging Large Language Models for Accurate Retrieval of Patient Information From Medical Reports: Systematic Evaluation Study.* Jmir ai, 2025. **4**: p. e68776.
14. Rastogi, C., et al., *How do authors' perceptions of their papers compare with co-authors' perceptions and peer-review decisions?* PLoS One, 2024. **19**(4): p. e0300710.
15. Paci, W., A. Panunzi, and S. Pezzelle. *They want to pretend not to understand: The Limits of Current LLMs in Interpreting Implicit Content of Political Discourse*. in *Findings of the Association for Computational Linguistics: ACL 2025*. 2025.
16. Ntinopoulos, V., et al., *Large language models for data extraction from unstructured and semi-structured electronic health records: a multiple model performance evaluation.* BMJ Health & Care Informatics, 2025. **32**(1): p. e101139.
17. Li, Z., et al., *RAPID: Reliable and efficient Automatic generation of submission rePortIng checklists with large language moDels.* Journal of the American Medical Informatics Association, 2025. **32**(8): p. 1340-1349.

**SUPPLEMENTARY MATERIAL**

**Supplementary Text-1:**
**List of Included Articles.** Complete references for the 17 observational studies assessed for the STROBE checklist agreement

1.  Larid, G., et al., *Detection of hypophosphatasia in hospitalised adults in rheumatology and internal medicine departments: a multicentre study over 10 years.* RMD Open, 2024. **10**(2).
2.  Bastard, L., et al., *Risk of serious infection associated with different classes of targeted therapies used in psoriatic arthritis: a nationwide cohort study from the French Health Insurance Database (SNDS).* RMD Open, 2024. **10**(1).
3.  Sloan, M., et al., *Prevalence and identification of neuropsychiatric symptoms in systemic autoimmune rheumatic diseases: an international mixed methods study.* Rheumatology (Oxford), 2024. **63**(5): p. 1259-1272.
4.  Crossfield, S.S.R., et al., *Changes in the pharmacological management of rheumatoid arthritis over two decades.* Rheumatology (Oxford), 2021. **60**(9): p. 4141-4151.
5.  Spierings, J., et al., *How do patients with systemic sclerosis experience currently provided healthcare and how should we measure its quality?* Rheumatology (Oxford), 2020. **59**(6): p. 1226-1232.
6.  Garriga, C., et al., *Clinical and molecular associations with outcomes at 2 years after acute knee injury: a longitudinal study in the Knee Injury Cohort at the Kennedy (KICK).* Lancet Rheumatol, 2021. **3**(9): p. e648-e658.
7.  Jones, C.A., et al., *The effect of geographic location and payor type on provincial-wide delivery of the GLA:D program for hip and knee osteoarthritis in Alberta, Canada.* Osteoarthr Cartil Open, 2023. **5**(4): p. 100398.
8.  Mehta, P., et al., *Gastrointestinal manifestations in systemic lupus erythematosus: data from an Indian multi-institutional inception (INSPIRE) cohort.* Rheumatology (Oxford), 2025. **64**(1): p. 156-164.
9.  Mertens, M.G., et al., *Prediction models for treatment success after an interdisciplinary multimodal pain treatment program.* Semin Arthritis Rheum, 2025. **70**: p. 152592.
10. Palmowski, A., et al., *Initiation of glucocorticoids before entering rheumatology care associates with long-term glucocorticoid use in older adults with early rheumatoid arthritis: A joint analysis of Medicare and the Rheumatology Informatics System for Effectiveness (RISE) data.* Semin Arthritis Rheum, 2024. **68**: p. 152535.
11. van Straalen, J.W., et al., *Methotrexate therapy associated with a reduced rate of new-onset uveitis in patients with biological-naïve juvenile idiopathic arthritis.* RMD Open, 2023. **9**(2).
12. Li, H., et al., *Risk of gout flares after COVID-19 vaccination: A case-crossover study.* Semin Arthritis Rheum, 2022. **56**: p. 152059.
13. Rempenault, C., et al., *Risk of diverticulitis and gastrointestinal perforation in rheumatoid arthritis treated with tocilizumab compared to rituximab or abatacept.* Rheumatology (Oxford), 2022. **61**(3): p. 953-962.

14. Crossfield, S.S.R., et al., *Changes in ankylosing spondylitis incidence, prevalence and time to diagnosis over two decades.* RMD Open, 2021. **7**(3).
15. Chua, J., et al., *Stakeholders' preferences for osteoarthritis interventions in health services: A cross-sectional study using multi-criteria decision analysis.* Osteoarthr Cartil Open, 2020. **2**(4): p. 100110.
16. Van der Elst, K., et al., *One in five patients with rapidly and persistently controlled early rheumatoid arthritis report poor well-being after 1 year of treatment.* RMD Open, 2020. **6**(1).
17. Roth, E., et al., *Gastroesophageal reflux disease is associated with a more severe interstitial lung disease in systemic sclerosis in the EUSTAR cohort.* Rheumatology (Oxford), 2025. **64**(Si): p. Si63-si72.

**Supplementary Text-2:**

Standardized Prompting Protocol. This text details the complete, three-stage prompt sequence (including constraints, few-shot calibration examples, and target evaluation structure) administered to the language models to assess STROBE reporting adherence.

---

### #PROMPT-0 — Role, borders and logic of scoring

You are acting as an independent rater assessing adherence to the STROBE (Strengthening the Reporting of Observational Studies in Epidemiology) checklist.

Your task is to evaluate whether each STROBE item is reported in a manuscript, based strictly on the content of the provided PDF files.

Scoring options:
- Y (Yes): the item is clearly and explicitly reported
- N (No): the item is not reported or only partially reported
- NA (Not applicable): the item is inherently not applicable to the study design

Important rules:
- This assessment concerns reporting adherence only.
- Do NOT evaluate methodological quality, correctness, or risk of bias.
- If information is incomplete, implicit, vague, or only partially reported, it must be scored as "N".
- Use "NA" only when the item is genuinely not applicable, not when information is missing.
- Do not infer or assume information that is not explicitly stated in the manuscript.

### #PROMPT-1- Calibration (2 articles and external responses)

You are provided with the following files:
- "STROBE.pdf": the current STROBE checklist to be used as the reference standard.
- "article1.pdf" and "article2.pdf": two example observational studies.
- "training_scores.xlsx": STROBE adherence scores for article1 and article2, completed by an external human evaluator who is not part of the main study raters.

These examples are provided solely to illustrate how the scoring rules were applied in practice and the level of explicitness required for assigning a "Yes" score.

They are NOT provided as gold-standard answers, and they must NOT be used for training, optimization, or learning beyond calibration of reporting thresholds.

Please review the STROBE checklist and examine how the scoring rules were applied to article1 and article2.

Do not generate any new scores at this stage.

Acknowledge once you are ready to apply the same scoring rules consistently to a new manuscript.

### #PROMPT-2 - Article Scoring

Now evaluate the attached manuscript using the same STROBE checklist and the same scoring rules applied previously.

Use the STROBE checklist provided in "STROBE.pdf".

Output requirements:
- Provide a table with exactly two columns:

1) STROBE item (including sub-items such as 1a, 1b)
2) Score (Y / N / NA)

Do not include explanations, justifications, page numbers, or comments.
Do not summarize or interpret the results.
Do not compare this manuscript to previous examples.
Produce a single, complete STROBE adherence table.

------------------------------------------------------------------------------------------------

**Supplementary Figure-1. Study selection flowchart.**

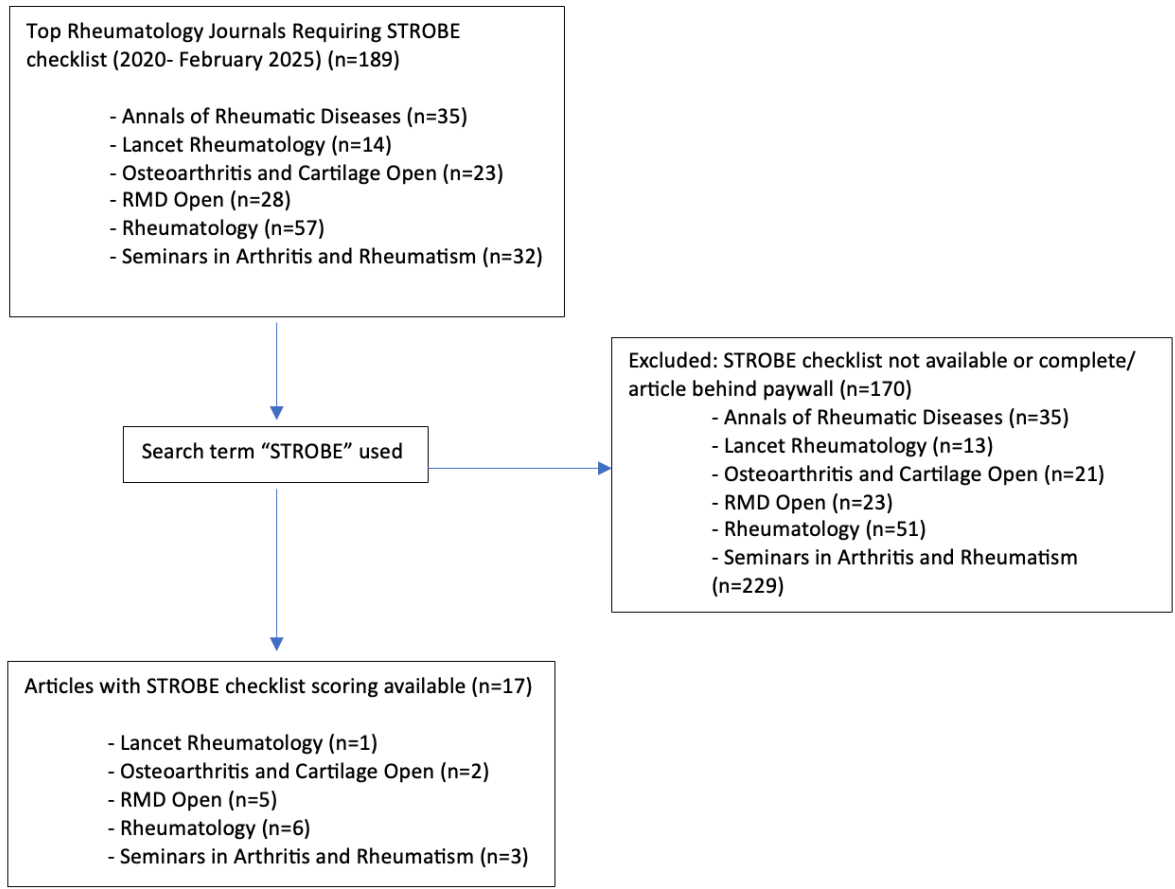

**Supplementary Figure-2: Item-specific inter-rater reliability across human reviewers and large language models.** Heatmaps display Gwet's AC1 agreement scores for each individual item and sub-item of the STROBE checklist. Each matrix compares the categorical reporting assessments (Yes / No / Not Applicable) among two large language models (ChatGPT-5.2 and Gemini 3 Pro), five independent human raters of varying clinical experience (Junior-1 to Senior), and the original study authors. The color gradient indicates the strength of agreement, ranging from 0.00 (poor agreement; light blue) to 1.00 (perfect agreement; dark blue).

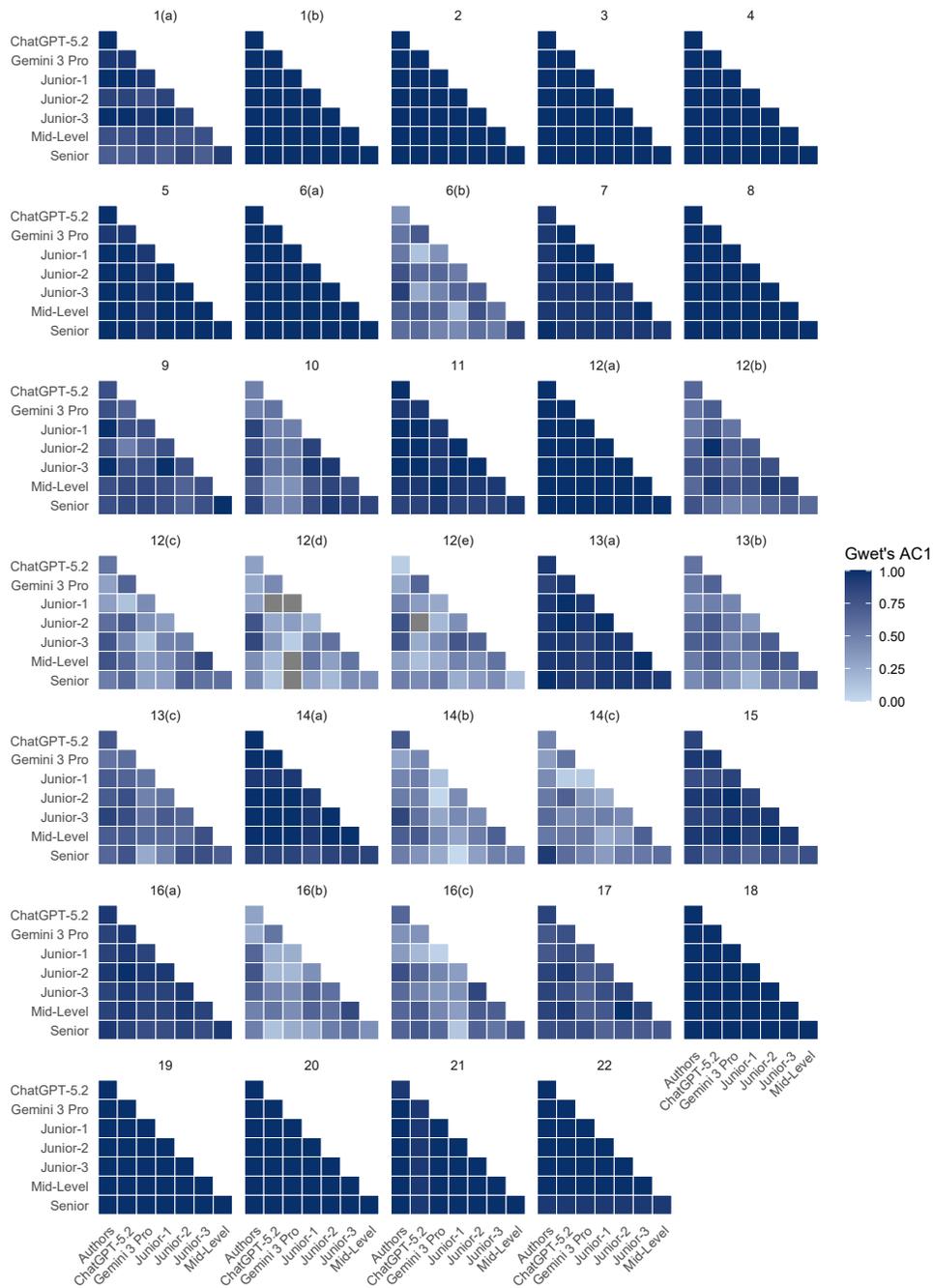